\documentclass[12pt]{article}
\pdfoutput=1
\usepackage{amsmath,amssymb,amsfonts,amsthm,bm,bbm,cancel,wasysym}
\usepackage{epsfig,graphics,graphicx,epstopdf,caption,subcaption}
\usepackage[numbers,sort&compress]{natbib}
\graphicspath{{Charts/}}
\usepackage{array,booktabs,colortbl,colordvi,multirow}
\usepackage{colordvi,color,xcolor}
\usepackage{hyperref}
\usepackage{rotating}
\usepackage{comment}
\usepackage{feynmf}

\usepackage[symbol]{footmisc}
\renewcommand{\thefootnote}{\fnsymbol{footnote}}

\begin{document}

\begin{center}

{\Large {\bf Gravitational Dark Matter from Minimal Preheating}}\\

\vspace*{0.75cm}

{Ruopeng Zhang and Sibo Zheng}

\vspace{0.5cm}
{Department of Physics, Chongqing University, Chongqing 401331, China }

\end{center}
\vspace{.5cm}

\begin{abstract}
\noindent 
Following our previous work, we continue to explore gravitational dark matter production during the minimal preheating caused by inflaton self-resonance.  In this situation there is only one dimensionless index parameter $n$ characterizing the inflation potential after the end of inflation, 
which leads to a robust prediction on the gravitational dark matter relic abundance. 
Using lattice method to handle the non-perturbative evolutions of relevant quantities during the inflaton self-resonance,
we derive the gravitational dark matter relic abundance arising from both the inflaton condensate and fluctuation annihilation.
While being absent for $n=2$, the former one can instead dominate over the later one for $n=4,6$.
Our results show that gravitational dark matter mass of $1.04~(2.66)\times 10^{14}$ GeV accommodates the observed value of dark matter relic abundance for $n=4$ (6).
\end{abstract}

\renewcommand{\thefootnote}{\arabic{footnote}}
\setcounter{footnote}{0}
\thispagestyle{empty}
\vfill
\newpage
\setcounter{page}{1}

\tableofcontents

\section{Introduction}
\label{int}
In the standard cosmological model, established by the observations of cosmic microwave background and large-scale structure, 
an epoch is believed to take place between the inflationary and the radiation-dominated Universe. 
During this epoch, there is a mechanism to convert the inflaton energy density after the end of inflation into the radiation energy density.
Depending on the mechanism being either perturbative or non-perturbative, this epoch is called as reheating \cite{Bassett:2005xm} or preheating \cite{Amin:2014eta} respectively.  
Alongside the transfer of the inflaton energy density to the radiation energy density,
in principle dark matter (DM) - the unknown major matter content of Universe - can be also partially or entirely produced during this epoch.

DM production during (p)reheating has been studied in depth either via a hypothetical inflaton-DM interaction \cite{Bernal:2018hjm,Lebedev:2021tas,Garcia:2021iag,Garcia:2022vwm} or purely gravitational portal \cite{Markkanen:2015xuw,Fairbairn:2018bsw,Hashiba:2018tbu, Ema:2019yrd,Ahmed:2020fhc,Babichev:2020yeo,Karam:2020rpa,Ling:2021zlj,Mambrini:2021zpp,Barman:2021ugy,Clery:2021bwz,Haque:2021mab,Kaneta:2022gug,Ahmed:2022tfm,Klaric:2022qly,Barman:2022qgt,Lebedev:2022vwf, Zhang:2023xcd}. 
In practice, for any DM candidate the gravitational-portal contribution is intrinsic, 
which is increasingly strengthened by null results of experiments aiming to detect non-gravitational DM. 
Very recently, we have reported a viability of gravitational dark matter (GDM) production in Higgs preheating \cite{Zhang:2023xcd},
where parameter resonance triggered by a quartic interaction between the inflaton and SM Higgs leads to an explosive production of relativistic Higgs particles with high energies above $\omega_{*}$ - a characteristic mass scale of the preheating model.
The Higgs particles can subsequently annihilate into GDM particles through the gravitational portal. 
Our work has demonstrated a critical point that preheating is able to  provide a circumstance under which the GDM relic abundance is accurately produced. 

In this work we continue to  investigate the GDM production in the minimal preheating caused by inflaton self-resonance \cite{Amin:2010dc,Amin:2011hj,Lozanov:2016hid,Lozanov:2017hjm}.
Unlike in the Higgs preheating, there is no need of hypothetical interaction in the minimal preheating, 
 except with only a dimensionless parameter $n$ appearing in the inflation potential after the end of inflation, i.e, $V(\phi)\sim \phi^{n}$.
The aforementioned studies have shown that the inflaton self-resonance happens to occur for the specifical values of $n=4$ and 6,
which yields an explosive production of inflaton fluctuations.
To carry out the main task of this work, we firstly utilize a lattice method to handle the quantitate analysis of preheating due to the inflaton self-resonance. While the results about the non-perturbative evolutions of relevant energy densities and the phase-space distribution function (PDF) of inflaton flucutations have been present in the literature \cite{Garcia:2023dyf},
they have to be reproduced in order to calculate the GDM relic abundance.
We then analyze the GDM production, arising from both the inflaton condensate and fluctuation annihilations.
While being absent for $n=2$, the inflaton condensate annihilation contribution in the GDM mass range of $m_{\chi}>\omega_{*}$  reopens for $n=4,6$.

The structure of the paper is organized as follows.
In Sec.\ref{mp} we show the main features of inflaton self-resonance for $n=4$ and 6 in terms of a lattice treatment.
The obtained results about the evolutions of relevant energy densities and the PDFs of inflaton fluctuations
will be applied to derive the GDM relic abundance. 
In Sec.\ref{gdm} we discuss both the inflaton condensate and fluctuation annihilation contribution to the GDM production,
where the former is verified to dominate over the later in the GDM mass range of $m_{\chi}\sim \omega_{*}$.
Our results show that GDM mass of $1.04~(2.66)\times 10^{14}$ GeV saturates the observed value of DM relic abundance \cite{Planck:2018jri} for $n=4$ (6).
Since this result does not depend on any other model parameters rather than the index $n$, it can be understood as a robust prediction of the minimal preheating due to the inflaton self-resonance. 
Finally, we conclude in Sec.\ref{con}. 
In appendix.\ref{If}, we present the details of inflaton fluctuation annihilation contribution to the GDM relic abundance. 

Note: throughout the text we take reduced Planck mass $M_{P}=2.4\times 10^{18}$ GeV, the subscript ``end" denotes the end of inflation, 
and the GDM mass (time $t^{-1}$ ) is in units of $\omega_{*}$ to be specified.

\section{The minimal preheating}
\label{mp}
To set the stage of early Universe after inflation, 
in this work we use the $\alpha$-attractor T-model of inflation \cite{Kallosh:2013hoa, Kallosh:2013yoa} with the inflation potential 
\begin{eqnarray}{\label{V}}
V(\phi)&=&\lambda M^{4}_{P}\left[\sqrt{6}\tanh\left(\frac{\phi}{\sqrt{6}M_P}\right)\right]^{n}\nonumber\\
&=&\left\{
\begin{array}{lcl}
\lambda M^{4}_{P}, ~~~~~~~~~~\mid\phi\mid>>M_{P},\\
\lambda\frac{\phi^{n}}{M^{n-4}_P}, ~~~~~~~~\mid\phi\mid<<M_{P},\\
\end{array}\right.
\end{eqnarray}
where $\lambda$ is a dimensionless parameter, $\phi$ is the inflaton scalar field, and $n$ is the index chosen as an even number in order to keep the positivity of $V(\phi)$ after inflation. 

After the end of inflation, $\phi$ begins to oscillate around the minimal value of $V(\phi)$ as shown in eq.(\ref{V}).
For such an oscillation the initial conditions \cite{Garcia:2020wiy} can be determined by Planck 2018 data \cite{Planck:2018jri} and the adopted criteria - the inflation ends at $\ddot{a}=0$ - which corresponds to
\begin{eqnarray}{\label{inic}}
\phi_{\rm{end}}=\sqrt{\frac{3}{8}}M_{P}\rm{ln}\left[\frac{1}{2}+\frac{n}{3}(n+\sqrt{n^{2}+3})\right],
\end{eqnarray}
with $a$ the scale factor.
Table \ref{ini} collects the initial conditions with respect to an explicit value of $n$.

\begin{table}
\begin{center}
\begin{tabular}{ccc}
\hline\hline
$\rm{n}$~~~ & $\lambda$~~~~ &  $\phi_{\rm{end}}/M_{P}$ \\ \hline
2~~~ & $2\times 10^{-11}$ &  $0.78$ \\
4~~~ &  $3.4\times 10^{-12}$& $1.50$ \\
6~~~ & $5.7\times 10^{-13}$  & $1.97$\\
\hline \hline
\end{tabular}
\caption{The initial conditions after the end of inflation.}
\label{ini}
\end{center}
\end{table}

For certain values of $n$ in Table \ref{ini} there exists an explosive production of the inflaton fluctuations due to the inflaton self-resonance \cite{Amin:2010dc,Amin:2011hj,Lozanov:2016hid,Lozanov:2017hjm}.
The aim of the rest of this section is to verify the non-perturbative phenomena of inflaton self-resonance, 
and to obtain the evolutions of relevant energy densities and the PDFs of inflaton fluctuations for the later use in Sec.\ref{gdm}.

\subsection{Lattice treatment}
We start with the equation of motion for the inflaton fluctuation $\delta\phi$
\begin{eqnarray}{\label{eom}}
\ddot{\delta\phi}+3H\dot{\delta\phi}-\frac{\nabla^{2}}{a^{2}}\delta\phi+\frac{\partial^{2}V}{\partial \phi^{2}}\delta\phi=0,
\end{eqnarray}
where $H$ is the Hubble parameter and $\nabla$ is the derivative over space coordinates. 
Rescaling the fluctuation as $X=a\delta\phi$, switching to the conformal time $\tau$, and expanding the quantized $X$ in momentum space as 
\begin{eqnarray}{\label{Xexp}}
X(\tau,\mathbf{x})=\int \frac{d^{3}\mathbf{k}}{(2\pi)^{3/2}}e^{-i\mathbf{k}\cdot\mathbf{x}}[X_{k}(\tau)a_{\mathbf{k}}+X^{*}_{k}(\tau)a^{\dag}_{\mathbf{k}}],
\end{eqnarray}
with $\mathbf{k}$ the comoving momentum, eq.(\ref{eom}) is rewritten as 
\begin{eqnarray}{\label{eomn}}
X''_{k}+\omega^{2}_{k}X_{k}=0,
\end{eqnarray}
where $' $ denotes the derivative with respect to conformal time and 
\begin{eqnarray}{\label{coef}}
\omega^{2}_{k}=k^{2}+a^{2}\frac{\partial^{2}V}{\partial \phi^{2}},
\end{eqnarray}
with $k=\mid \mathbf{k}\mid$.

In terms of publicly available lattice code CosmoLattice \cite{Figueroa:2020rrl, Figueroa:2021yhd} 
we solve eq.(\ref{eomn}) with the initial conditions of positive-frequency Bunch-Davies vacuum:
\begin{eqnarray}{\label{BD}}
X_{k}(\tau_{0})=\frac{1}{\sqrt{2\omega_{k}}},~~~~~X'_{k}(\tau_{0})=-\frac{i\omega_{k}}{\sqrt{2\omega_{k}}}.
\end{eqnarray}
The lattice calculation provides us the PDF of $\delta\phi$ via occupation number $n_k$ \cite{Kofman:1997yn} 
\begin{eqnarray}{\label{PDFd}}
f_{\delta\phi}=n_{k}=\frac{1}{2\omega_{k}}\mid\omega_{k}X_{k}-iX'_{k}\mid^{2}.
\end{eqnarray}
Using eq.(\ref{PDFd}) one finds the number density and energy density of the fluctuations
\begin{eqnarray}{\label{density}}
n_{\delta\phi}&=& \frac{1}{(2\pi)^{3}}\left(\frac{a_{\rm{end}}}{a}\right)^{3}\int d^{3}\mathbf{k}~n_{k},  \nonumber\\
\rho_{\delta\phi}&=&\frac{1}{(2\pi)^{3}}\left(\frac{a_{\rm{end}}}{a}\right)^{4}\int d^{3}\mathbf{k}~n_{k}\omega_{k}.
\end{eqnarray}

\begin{figure}
\centering
\includegraphics[width=8cm,height=8cm]{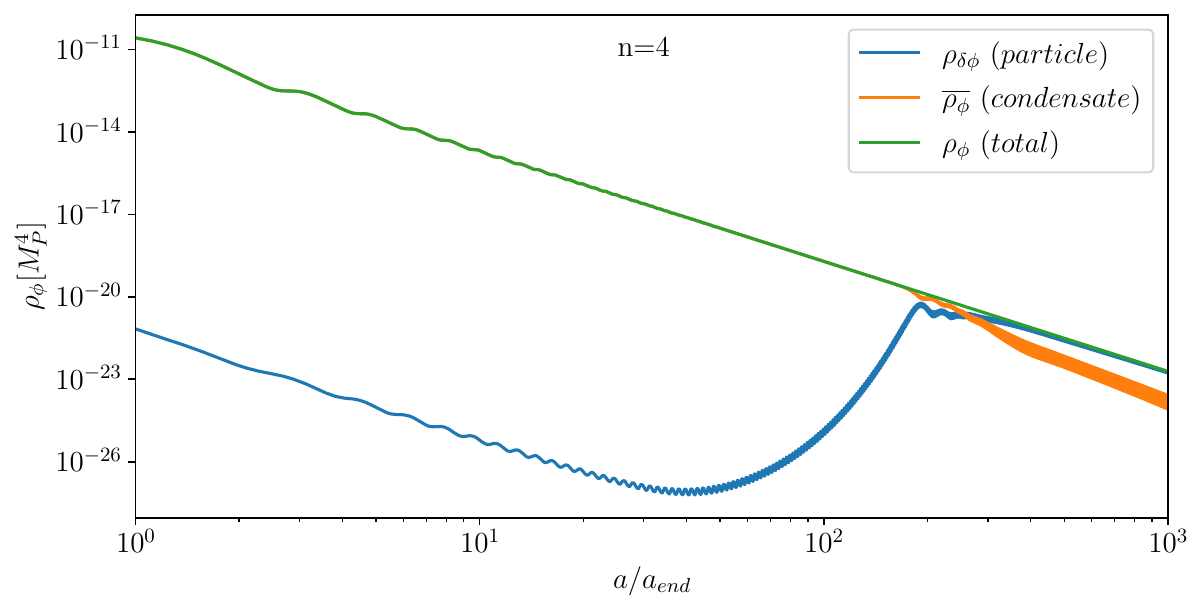}
\includegraphics[width=8cm,height=8cm]{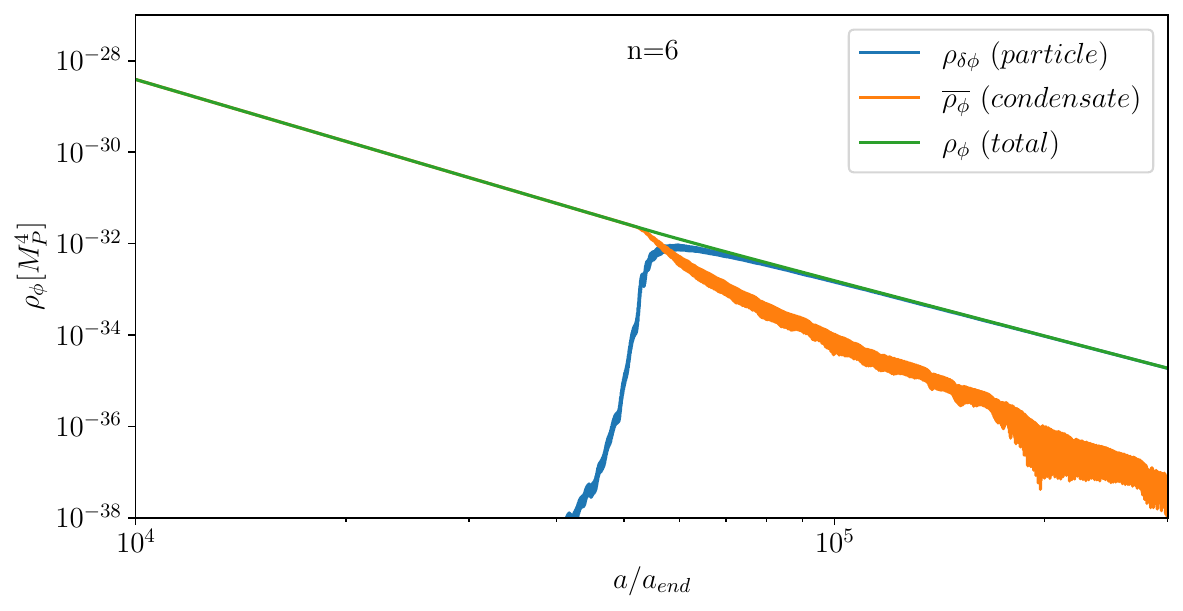}
\centering
\caption{Left: the evolutions of inflaton condensate energy density $\bar{\rho}_{\phi}$ and inflaton fluctuation energy density $\rho_{\delta\phi}$  for $n=4$. 
Right: the same as the left panel but for $n=6$.}
\label{rho}
\end{figure}

Along with a verification of the absence of inflaton self-resonance for $n=2$, 
fig.\ref{rho} shows the evolutions of inflaton fluctuation energy density $\rho_{\delta\phi}$ defined by eq.(\ref{density}), 
inflaton condensate energy density averaged over a period of oscillation time,
and total energy density $\rho_{\phi}=\bar{\rho}_{\phi}+\rho_{\delta\phi}$ for $n=4$ (left) and $n=6$ (right) respectively.
In this figure our lattice results agree well with those of \cite{Garcia:2023dyf},
where the phenomena of self-resonance is clearly seen, 
as the inflaton fluctuations are explosively produced at the time with respect to $a/a_{\rm{end}}\sim 2\times10^{2}$ and $\sim 5\times 10^{4}$ for $n=4$ and $n=6$, respectively. 
Compared to the case of $n=4$, the inflaton self-resonance in the case of $n=6$ occurs at a relatively later time,
which can be understood as a result of the different dependences of $\omega_k$ on $a$ between these two cases.
As seen in eq.(\ref{coef}), $\omega_{k}\sim a^{0}$ ($a^{-1/2}$) for $n=4$ (6),
implying a relatively postponed growth of the inflaton fluctuations for $n=6$.

Apart from in the evolutions of the above energy densities, 
the self-resonance can be also seen in the PDF of inflaton fluctuations. 
For illustration, the PDF for $n=4$ is shown in the left panel of fig.\ref{PDF} in the Appendix,
where the occupation number peaks near $k/\omega_{*}\approx 2$ consistent with \cite{Garcia:2023eol}.
Note, in this figure the comoving momentum is in units of $\omega_{*}$,
which is defined as $\omega_{*}=\sqrt{\phi^{-1}(\partial V/\partial \phi)}\mid_{a=a_{\rm{end}}}$  \cite{Figueroa:2020rrl, Figueroa:2021yhd}. 
According to the initial conditions in Table \ref{ini}, $\omega_{*}=1.37~(1.77)\times 10^{13}$ GeV for $n=4$ (6).
In the appendix, the PDFs of inflaton fluctuations are applied to estimate the inflaton fluctuation annihilation contribution to the GDM relic abundance discussed below.

\section{Gravitational dark matter}
\label{gdm}
Having gained the background materials of minimal preheating induced by the inflaton self-resonance, 
we now derive the GDM relic abundance in terms of the Boltzmann equation\footnote{In eq.(\ref{Boltn}) an additional SM annihilation rate $R_{\rm{SM}}$  \cite{Garny:2015sjg,Tang:2017hvq,Garny:2017kha,Chianese:2020yjo}  also appears during reheating that follows the preheating. 
Since $R_{\rm{SM}}\sim T^{8}_{\rm{r}}/M^{4}_{P}$ with $T_{\rm{r}}$ the reheating temperature,
the relative importance of $R_{\rm{SM}}$ to $R_{\phi}$ and $R_{\delta\phi}$ depends on the value of $T_{\rm{r}}$.
Consider that $T_{\rm{r}}$ relies on how the inflaton fluctuation energy density is converted into the radiation energy density, which is beyond the scope of this study, we simply assume $T_{\rm{r}}<<m_{\chi}$, under which the $R_{\rm{SM}}$ contribution can be safely neglected.
On the contrary, if $m_{\chi}<T_{\rm{r}}$, the $R_{\rm{SM}}$ rate should be taken into account.} 
\begin{eqnarray}\label{Boltn}
\dot{n}_{\chi}+3Hn_{\chi}=R_{\delta\phi}+R_{\phi},
\end{eqnarray}
which is valid despite the evolutions of $\rho_{\phi}$ and $\rho_{\delta\phi}$ being non-perturbative.
Here, $R_{\phi}$ and  $R_{\delta\phi}$ represent the inflaton condensate annihilation and the inflaton fluctuation annihilation rate respectively.

Both the rates $\rho_{\phi}$ and $\rho_{\delta\phi}$ are calculated in terms of quantum field theory techniques based on the gravitational interaction 
\begin{eqnarray}{\label{int}}
\mathcal{L}_{\rm{int}}=-\frac{h^{\mu\nu}}{M_{P}}\left[T_{\mu\nu}^{\phi}+T_{\mu\nu}^{\rm{SM}}+T_{\mu\nu}^{\rm{\chi}}\right],
\end{eqnarray}
where $h_{\mu\nu}$ is the gravitational field and $T^{(j)}_{\mu\nu}$ are the energy-momentum tensors of sectors $j=\{\rm{SM}, \phi, \chi\}$.
In this work we explicitly consider the GDM $\chi$ being a Dirac fermion. An extension to scalar- or vector-like GDM is straightforward, which is beyond the scope of the present work.

\subsection{Gravitational annihilation rates}
The rate $R_{\phi}$ in eq.(\ref{Boltn})  due to the inflaton condensate annihilation $\phi\phi\rightarrow \chi\chi$ is given by \cite{Clery:2021bwz, Ahmed:2022tfm}
\begin{eqnarray}\label{Ra}
R_{\phi}(t)=\frac{\rho^{2}_{\phi}}{2\pi M^{4}_{P}}\sum_{\ell=1}^{\infty}\mid\mathcal{P}^{n}_{\ell}\mid^{2}\frac{m^{2}_{\chi}}{(\ell\omega)^{2}}\sqrt{1-\frac{4m^{2}_{\chi}}{(\ell\omega)^{2}}}, 
\end{eqnarray}
by following the fact that $\phi$ is an excitation out of vacuum rather than a real particle.
Here, $\mathcal{P}^{n}_{\ell}$ are the Fourier coefficients
\begin{eqnarray}\label{Fourc}
\mathcal{P}^{n}_{\ell}=\frac{1}{\mathcal{T}}\int^{\mathcal{T}}_{0}dt\mathcal{P}^{n}(t)e^{i\ell\omega t},
\end{eqnarray}
of fast-oscillating function \cite{Ahmed:2022tfm}
\begin{eqnarray}\label{Ps}
\mathcal{P}^{n}(t)\approx \mathcal{I}^{-1}_{z}(\frac{1}{n},\frac{1}{2}),
\end{eqnarray}
where $\mathcal{I}^{-1}_{z}(i,j)$ the inverse of the regularized incomplete beta function,
\begin{eqnarray}\label{z}
z=1-\frac{2\sqrt{3}}{\pi}\frac{n+2}{n}\frac{\omega}{\sqrt{\lambda}M_P}\left(\frac{M_{P}}{\varphi_{\rm{end}}}\right)^{n/2}\left[\left(\frac{a}{a_{\rm{end}}}\right)^{\frac{3n}{n+2}}-1\right],
\end{eqnarray}
the oscillation frequency
\begin{eqnarray}\label{freq}
\omega=\sqrt{\frac{n^{2}(n-1)\pi\lambda}{2n-2}}\frac{\Gamma(\frac{n+2}{2n})}{\Gamma(\frac{1}{n})}M_{P}\left(\frac{\varphi(a)}{M_{P}}\right)^{\frac{n-2}{2}},
\end{eqnarray}
where the envelope function is given by
\begin{eqnarray}{\label{env}}
\varphi(a)\approx\varphi_{\rm{end}}\left(\frac{a_{\rm{end}}}{a}\right)^{\frac{6}{n+2}},
\end{eqnarray}
and finally $\mathcal{T}=2\pi/\omega$.

\begin{figure}
\centering
\includegraphics[width=8cm,height=8cm]{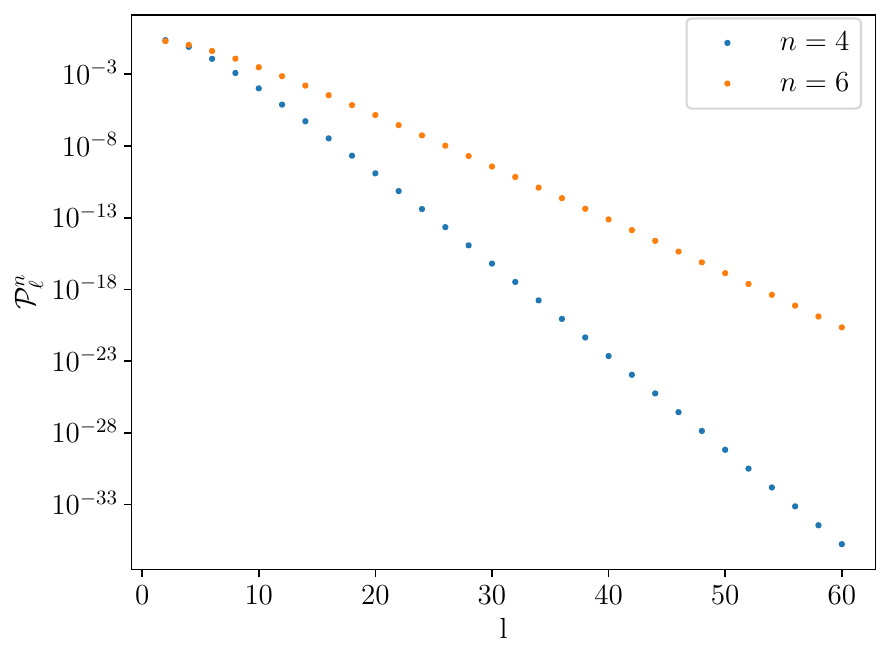}
\centering
\caption{The numerical value of $\mathcal{P}^{n}_{\ell}$ as function of $\ell$ for $n=4$ (in blue) and $n=6$ (in orange) respectively.}
\label{Pl}
\end{figure}

Unlike in the situation with $n=2$ where only the Fourier mode $\ell=2$ is non-zero,
the other even Fourier modes do not vanish in the cases of $n=4$ and 6.
This means that there is a sum over $\ell$ in eq.(\ref{Ra}).
Fortunately, the sum is greatly simplified by the following observation.
Given a value of $m_\chi$ for $n=4$ or 6, eq.(\ref{Ra})  implies $\ell\geq \ell_{\rm{min}}$ with
\begin{eqnarray}{\label{bound2}}
\ell_{\rm{min}}(a)=
 \left\{
\begin{array}{lcl}
\left[\frac{2m_{\chi}}{\omega(a)}\right]+1, ~~~~~\left[\frac{2m_{\chi}}{\omega(a)}\right]~\rm{odd}\\
\left[\frac{2m_{\chi}}{\omega(a)}\right]+2, ~~~~~\left[\frac{2m_{\chi}}{\omega(a)}\right]~\rm{even}\\
\end{array}\right.
\label{Lint}
\end{eqnarray}
As seen in eq.(\ref{freq}), $\omega\sim \omega_{*}a^{-1}$ and $\omega\sim \omega_{*}a^{-3/2}$ with respect to $n=4$ and $n=6$ respectively,
suggesting a higher $\ell_{\rm{min}}$ needed as $a$ increases.  
Thanks to a dramatically converge of $\mathcal{P}^{n}_{\ell}$  in the large $\ell$ regions as shown in fig.\ref{Pl},
it is suffice to take into account only the $\ell_{\rm{min}}$-mode contribution to $R_\phi$ with respect to an explicit $a$.

\begin{figure}
\centering
\includegraphics[width=8cm,height=8cm]{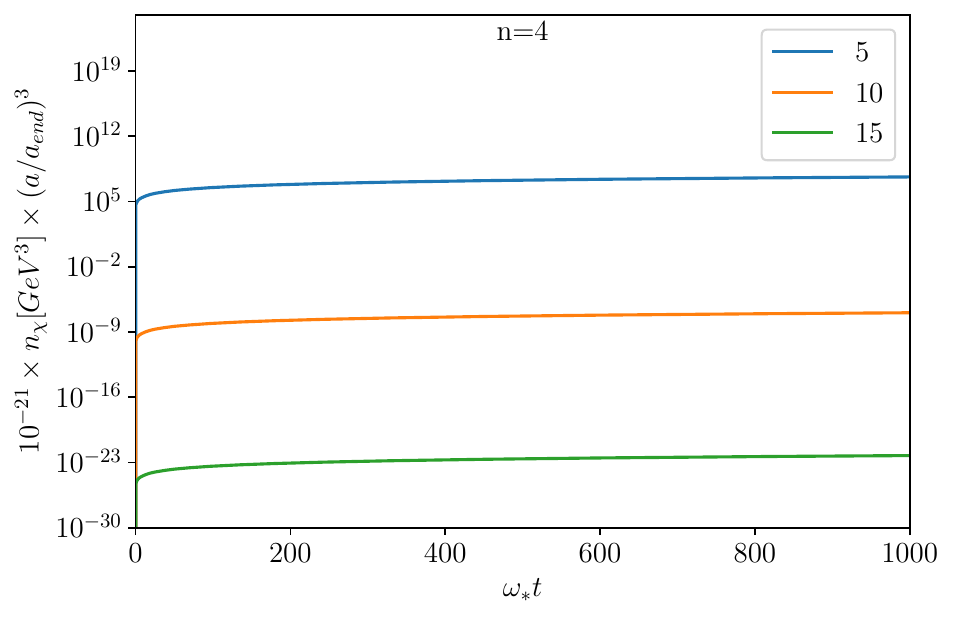}
\includegraphics[width=8cm,height=8cm]{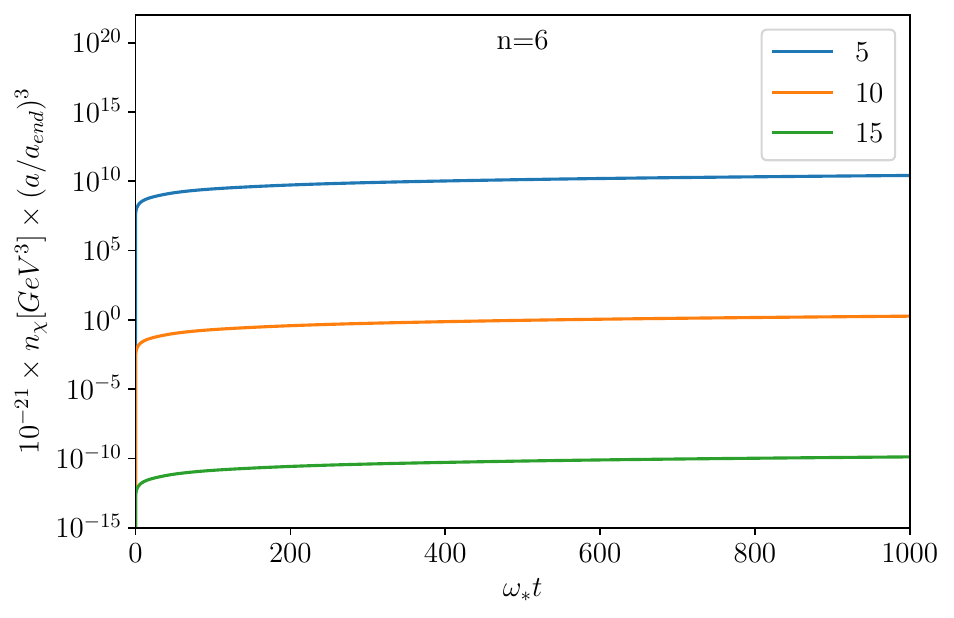}
\centering
\caption{The value of comoving number density $n_{\chi}(a/a_{\rm{end}})^{3}$ as function of time for various values of $m_{\chi}/\omega_{*}=\{5,10,15\}$ due to the $R_{\phi}$ contribution for $n=4$ (left) and $n=6$ (right) respectively.}
\label{Rphi}
\end{figure}

Fig.\ref{Rphi} shows the comoving number density $n_{\chi}a^{3}$ as function of time (in units of $\omega^{-1}_{*}$) for various values of GDM mass $m_{\chi}$ due to the $R_{\phi}$ contribution for $n=4$ (left) and $n=6$ (right) respectively, where the previous obtained evolutions of $\rho_{\phi}$ in fig.\ref{rho} have been used. 
In this figure the values of $n_{\chi}a^{3}$ is rapidly stabilized due to the rapid convergence of $\mathcal{P}^{n}_{\ell}$ in the large $\ell$ region as well as the rapid decrease of $\rho_{\phi}$ during preheating. 
As a result of a more slowly convergence of $\mathcal{P}^{6}_{\ell}$ relative to $\mathcal{P}^{4}_{\ell}$ as seen in fig.\ref{Pl}, 
given a fixed GDM mass the comoving number density for $n=6$ is several orders of magnitude larger than that for $n=4$.

In addition, comparing fig.\ref{Rphi} to the right panel of fig.\ref{PDF} in the Appendix.\ref{If},
one finds that the $R_{\phi}$ rate dominates over the $R_{\delta\phi}$ rate in the GDM mass range of $m_{\chi}\sim\omega_{*}$ for $n=4$ and 6.
In this appendix the $R_{\delta\phi}$ contribution for $n=4$ has been explicitly derived by applying the PDF (shown in the left panel of fig.\ref{PDF}) to the $R_{\delta\phi}$ rate for illustration.
There the comoving momentum range of $k$, excited by the self-resonance, is not large enough to let the $R_{\delta\phi}$ rate be comparable with the $R_{\phi}$ rate in the GDM mass range of $m_{\chi}\sim\omega_{*}$.
Because of this reason, we will neglect the $R_{\delta\phi}$ contribution to the GDM relic abundance in the following discussion.
This situation differs from that of parameter resonance in \cite{Zhang:2023xcd}, 
where although the $R_{\delta\phi}$ is absent in the GDM mass range of $m_{\chi}>\omega_{*}$ for $n=2$,
and the comoving momentum range of $k$ excited by the parameter resonance is able to provide a Higgs annihilation rate large enough to accommodate the observed DM relic abundance.

\subsection{Relic abundance}
Substituting the value of $n_{\chi}$ derived from eq.(\ref{Boltn}) into the following formula \cite{Zhang:2023xcd}
\begin{eqnarray}\label{rel}
\Omega_{\chi}h^{2}\approx1.6\times 10^{8}\cdot\left(\frac{g_{0}}{g_{\rm{r}}}\right)\frac{n_{\chi}}{\rho_{\delta\phi}^{3/4}}\left(\frac{m_{\chi}}{1~\rm{GeV}}\right),
\end{eqnarray}
one obtains the GDM relic abundance, where the values of $n_{\chi}$ and $\rho_{\delta\phi}$ are determined at the end of preheating with $g_{0}=43/11$ and $g_{\rm{r}}=427/4$. 
Note, the formula in eq.(\ref{rel}) is derived by recasting the result of \cite{Clery:2021bwz}.

\begin{figure}
\centering
\includegraphics[width=10cm,height=8cm]{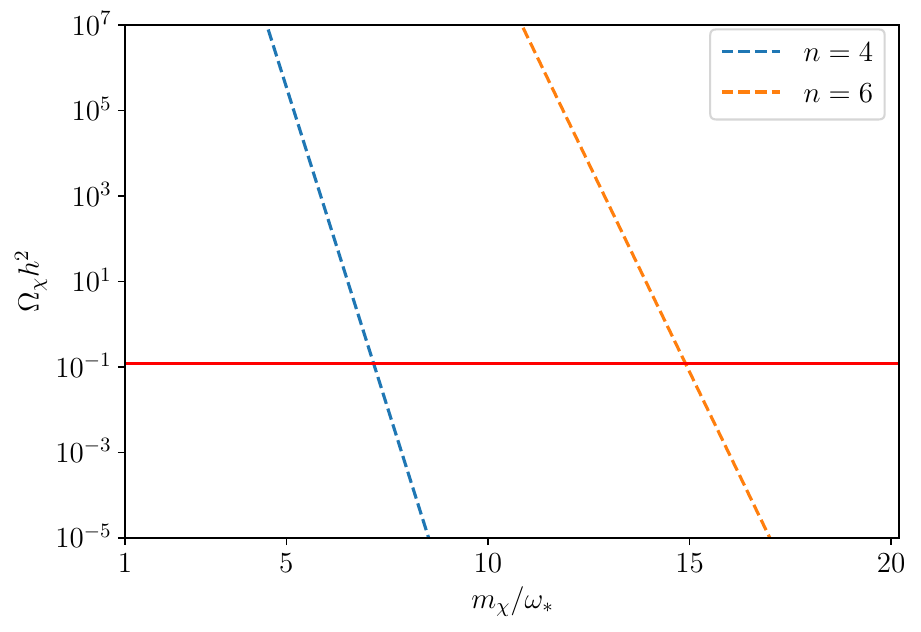}
\centering
\caption{The GDM relic abundance as function of $m_{\chi}$ due to the $R_{\phi}$ contribution for $n=4$  (in blue) and  $n=6$ (in orange) respectively,
where the red line corresponds to the value of $\Omega_{\chi}h^{2}=0.12$ and the two crossing points are not affected by the neglected $R_{\delta\phi}$ contribution.}
\label{relic}
\end{figure}

Fig.\ref{relic} shows the GDM relic abundance as function of $m_{\chi}$ for $n=4$ (in blue) and $n=6$ (in orange) respectively, 
either of which is mainly due to the $R_{\phi}$ contribution as explained above. 
Meanwhile, as a result of different convergences of $\mathcal{P}^{n}_{\ell}$ as previously mentioned, 
for an explicit $m_{\chi}$ the value of $\Omega_{\chi}h^{2}$ for $n=6$ is obviously larger than that for $n=4$.
This figure shows that while being overproduced in the GDM mass range of $m_{\chi}/\omega_{*}\leq 1$ the GDM relic abundance is able to accommodate the observed value of DM relic abundance $\Omega_{\chi}h^{2}\approx 0.12$ (in red)
for $m_{\chi}/\omega_{*}\sim 7.6~(15.0)$ for $n=4$ (6). 
These two crossing points correspond to the values of $m_{\chi}\approx 1.04~( 2.66)\times 10^{14}$ GeV, 
which can be understood as the robust prediction\footnote{The main uncertainty arises from the definition about the end of preheating in eq.(\ref{rel}). Here, we adopt the criteria that the preheating ends at the time with respect to $\rho_{\delta\phi}=\bar{\rho}_{\phi}$.}
 of minimal preheating.
Because they do not depend on any other model parameters except the index $n$.

\newpage
\section{Conclusion}
\label{con}

Motived by our previous study of GDM due to the Higgs preheating, 
in this work we have explored the GDM production in the minimal preheating triggered by the inflaton self-resonance.
Unlike in the Higgs preheating caused by the parameter resonance, 
there is no need of additional model parameters in the self-resonance except the index parameter $n$ fixed to be $4$ or 6,
which therefore results in a robust prediction on the GDM relic abundance.
After using the lattice method to handle the non-perturbative evolutions of relevant energy densities and the PDFs of inflaton fluctuations during the minimal preheating,
and applying them to calculate the inflaton condensate and fluctuation annihilation contribution to the GDM relic abundance, 
we show that GDM mass of $\sim 1.04~( 2.66)\times 10^{14}$ GeV accommodates the observed value of DM relic abundance for $n=4$ (6).

While being correctly produced in the early Universe, the GDM is a lack of effective probes. 
From a phenomenological viewpoint, the GDM can be understood as a special freeze-in DM with the feeble coupling being gravity.
Nevertheless, few methods developed to test freeze-in DM based on DM decay, annihilation, or scattering off SM particles, are however not viable for the GDM,
as the GDM is absolutely stable and the GDM scattering and annihilation cross sections are always Planck-scale suppressed. 
Looking for non-Planck-scale suppressed measurements such as equation of state of DM, which can be extracted from future observations of large-scale structure,  may be more promising GDM probes. 
This topic is left for a future study.

\appendix
\numberwithin{equation}{section}

\section{Inflaton fluctuation annihilation}
\label{If}
In this appendix we present the inflaton fluctuation annihilation contribution to the GDM relic abundance,
which gives rise to the rate $R_{\delta\phi}$ in eq.(\ref{Boltn}) as \cite{Clery:2021bwz}  
\begin{eqnarray}\label{Rb}
R_{\delta\phi}(t)\approx g_{\chi} \int d\Pi_{1}d\Pi_{2}d\Pi_{3}d\Pi_{4}(2\pi)^{4}\delta^{(4)}(p_{1}+p_{2}-p_{3}-p_{4})\mid\mathcal{M}\mid_{12\rightarrow 34}^{2}f_{1}f_{2},
\end{eqnarray}
where $d\Pi_{i}=d^{3}\mathbf{p}_{i}/((2\pi)^{3}(2p^{0}_{i}))$ are the phase space elements with $\mid\mathbf{p}_{i}\mid$ the physical momenta and $i=1-4$, 
$f_{j}$ is the out-of-equilibrium PDF of inflaton fluctuations with $j=1,2$, 
$g_\chi$ is the number of degrees of freedom for the DM, the angles $\theta_{13}$ and $\theta_{12}$ are given by
\begin{eqnarray}\label{angles}
s&=&2p^{0}_{1}p^{0}_{2}(1-\cos\theta_{12}), \nonumber\\
t&=&s(\cos\theta_{13}-1)/2,
\end{eqnarray}
with $s$ and $t$ the two Mandelstam variables,
and finally $\mid\mathcal{M}\mid_{12\rightarrow 34}^{2}$ is the transition amplitude given by
\begin{eqnarray}\label{amp}
\mid\mathcal{M}\mid_{12\rightarrow 34}^{2}&=&\frac{-t(s+t)(s+2t)^{2}}{4M^{4}_{P}s^{2}}.
\end{eqnarray}

To directly use the PDF of inflaton fluctuations, 
we have to transform the physical momenta $p_{i}=\mid\mathbf{p}_{i}\mid$ to the comoving momenta $k_i$ via $p_{i}=k_{i}(a_{\rm{end}}/a)$. 
Accordingly, the boundary condition of momentum integration in eq.(\ref{Rb}) becomes  
\begin{eqnarray}\label{bound1}
\left(k_{1}+k_{2}\right)\left(\frac{a_{\rm{end}}}{a}\right)\geq 2m_{\chi}.
\end{eqnarray}
Eq.(\ref{bound1}) suggests that for a fixed value of $m_{\chi}$ a larger value of $(k_{1}+k_{2})$ is required as $a$ increases,
which kinetically prohibits the GDM production as the preheating proceeds.

\begin{figure}
\centering
\includegraphics[width=8cm,height=8cm]{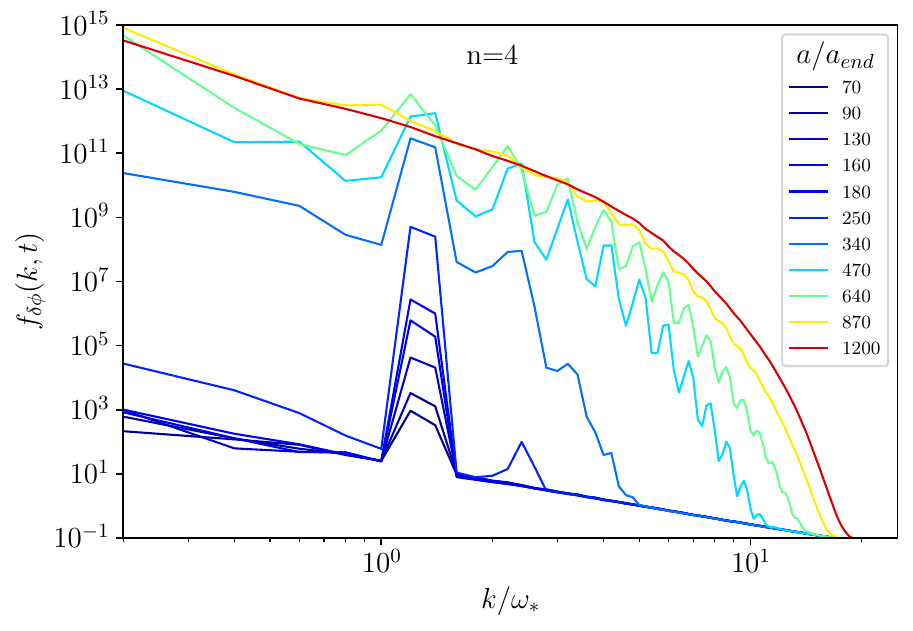}
\includegraphics[width=8cm,height=8cm]{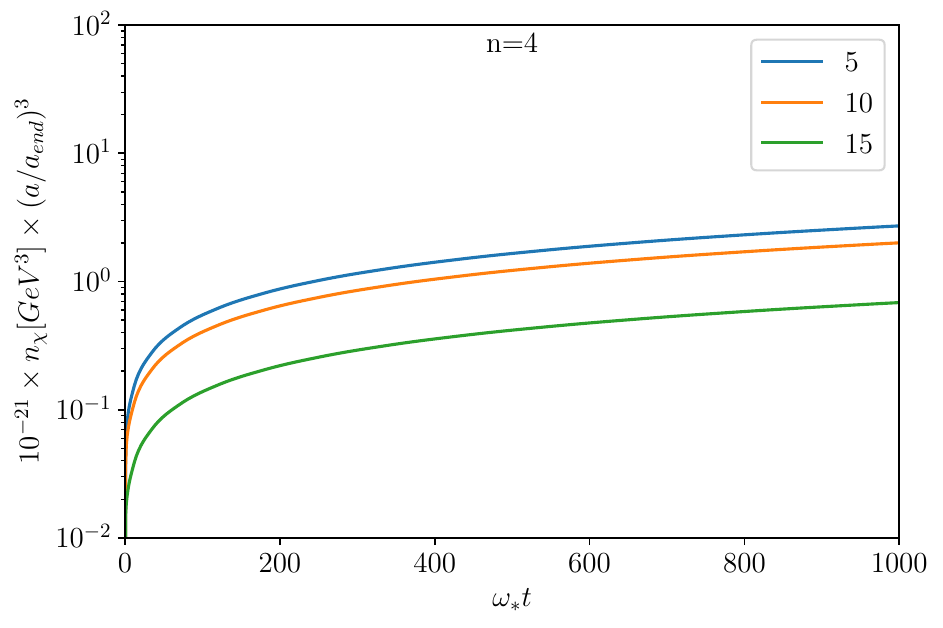}
\centering
\caption{An illustration of the inflaton fluctuations for $n=4$. Left: the PDF of inflaton fluctuations. Right: the inflaton fluctuation annihilation contribution to the comoving number density for various values of $m_{\chi}/\omega_{*}=\{5,10,15\}$.}
\label{PDF}
\end{figure}

For illustration, we show in fig.\ref{PDF} the PDF (left) where $k$ is in units of $\omega_{*}$,
and the $R_{\delta\phi}$ contribution to the comoving number density $n_{\chi}a^{3}$ (right) by substituting the PDF into eq.(\ref{Rb}) for $n=4$.
In the left panel, the lattice result agrees well with that of \cite{Garcia:2023eol}. 
In the right panel, the value of $n_{\chi}(a/a_{\rm{end}})^{3}$ peaks around $m_{\chi}/\omega_{*}\sim 10$, 
which gives rise to the maximal value of $\Omega_{\chi}h^{2}\approx 4\times10^{-3}$.
This implies that the $R_{\delta\phi}$ contribution due to the self-resonance is insufficient to accommodate the observed value of $\Omega_{\chi}h^{2}$.
On the contrary, the Higgs annihilation contribution to  $\Omega_{\chi}h^{2}$ due to the parameter resonance in \cite{Zhang:2023xcd} is large enough to address the required value of  $\Omega_{\chi}h^{2}$.

Compared to $n=4$, the constraint in eq.(\ref{bound1}) becomes more stringent for $n=6$. 
Therefore, the conclusion inferred from the $R_{\delta\phi}$ contribution to $\Omega_{\chi}h^{2}$ for $n=4$ in fig.\ref{PDF} also holds for $n=6$. 
 
\end{document}